\RequirePackage{ifpdf}
\ifpdf 
\documentclass[pdftex]{sigma}
\else
\documentclass{sigma}
\fi

\begin{document}
\allowdisplaybreaks

\renewcommand{\PaperNumber}{026}

\FirstPageHeading

\ShortArticleName{Conservation Laws of Discrete KdV Equation}

\ArticleName{Conservation Laws\\ of Discrete Korteweg--de Vries Equation}

\Author{Olexandr G. RASIN and Peter E. HYDON}
\AuthorNameForHeading{O.G. Rasin and P.E. Hydon}

\Address{Department of Mathematics and Statistics, University of Surrey,\\
Guildford, Surrey GU2 7XH, UK}
\Email{\href{mailto:o.rasin@surrey.ac.uk}{o.rasin@surrey.ac.uk},
 \href{mailto:p.hydon@surrey.ac.uk}{p.hydon@surrey.ac.uk}}

\ArticleDates{Received October 21, 2005, in final form December 06,
2005; Published online December 09, 2005}

\Abstract{All three-point and five-point conservation laws for the discrete Korteweg--de Vries equations are found.
These conservation laws satisfy a functional equation, which we solve by reducing it to a system
of partial differential equations. Our method uses computer algebra intensively, because the determining
functional equation is quite complicated.}


\Classification{70H33; 37K10; 39A05}

\section{Introduction}
\begin{figure}[th]
\centerline{\begin{minipage}{6.5cm}
\centerline{\includegraphics[height=4cm,width=4cm]{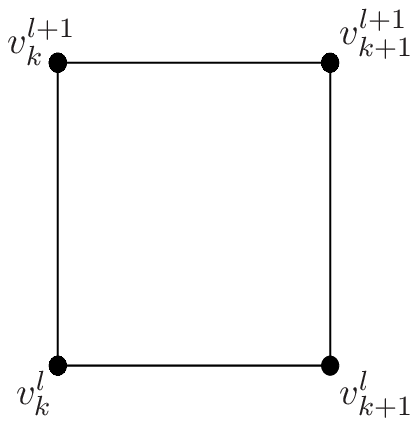}}
\vspace{-3mm}
\caption{Quad-graph.}\label{graph1} 
\end{minipage}
\quad
\begin{minipage}{6.5cm}
\centerline{\includegraphics[height=4cm,width=4cm]{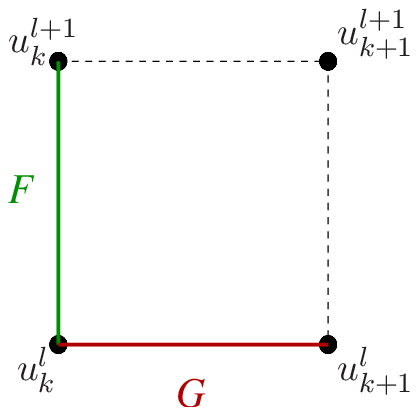}}
\vspace{-3mm}
\caption{Three-point conservation law.}
\label{graph2}
\end{minipage}}

\end{figure}

A direct method for calculation of conservation laws for partial difference equations
(P$\Delta$E's) was recently introduced by Hydon \cite{Hy0}. This method, which does not use Noether's Theorem,
has been used to calculate some low-order
conservation laws of various integrable difference equations that are defined on the quad-graph shown 
in Fig.~\ref{graph1}.
(For a classification of integrable quad-graph equations, see \cite{ABS}.)

In the current paper, we present a modified version of Hydon's direct method, and use it to derive conservation
laws of the discrete Korteweg--de Vries equation \cite{Hi1}:
\begin{gather}
\big(p+q+v_{k+1}^{l+1}-v_k^l\big)\big(q-p+v_{k+1}^l-v_k^{l+1}\big)=q^2-p^2,
\label{eqn0}
\end{gather}
which is an integrable quad-graph equation. Here $p,~q$ are parameters and $p^2\neq q^2$.
To simply matters, we use the transformation
\[
v_k^l=u_k^l\sqrt{q^2-p^2}-qk-pl
\]
to reduce (\ref{eqn0}) to
\[
\big(u_{k+1}^{l+1}-u_k^l\big)\big(u_{k+1}^l-u_k^{l+1}\big)=1.
\]
We shall call this equation dKdV. As with all integrable quad-graph equations, dKdV may be solved to
write any one of $u_k^l$, $u_{k+1}^l$, $u_k^{l+1}$, $u_{k+1}^{l+1}$ in terms of the other three. In particular,
we will write dKdV as either
\[
u_{k+1}^{l+1}=\omega, \qquad\text{where}\quad \omega=\frac{1}{u_{k+1}^l-u_k^{l+1}}+u_k^l,
\]
or
\[
 u_{k+1}^{l}=\Omega,\qquad\text{where}\quad
\Omega=\frac{1}{u_{k+1}^{l+1}-u_k^l}+u_k^{l+1}.
\]

A conservation law for a partial difference equation (P$\Delta$E) on $\mathbb{Z}^2$ is an expression of the form
\begin{gather*} 
(\mathbf S_k-\mathbf{ id})F+(\mathbf S_l-\mathbf{ id})G=0
\end{gather*}
that is satisfied on all solutions of the equation.
Here $\mathbf{id}$ is the identity mapping and $\mathbf S_k$, $\mathbf S_l$ are forward shifts of the
coordinates $k$ and $l$ respectively: 
\[ \mathbf
S_k:\big(k,l,u_k^l\big)\rightarrow\big(k+1,l,u_{k+1}^l\big),\qquad\mathbf S_l:\big(k,l,u_k^l\big)\rightarrow
\big(k,l+1,u_k^{l+1}\big). \]
A conservation law is trivial if it holds identically
(not just on solutions of the P$\Delta$E), or if $F$ and $G$ both vanish on all solutions of the equation.
We search for nontrivial conservation laws.

\section{Three-point conservation laws}
In this section we consider conservation laws that lie on the quad-graph. This means that the functions
$F$, $G$, $\mathbf S_kF$ and $\mathbf S_lG$ must depend upon
only $k$, $l$, $u_k^l$, $u_{k+1}^l$, $u_k^{l+1}$ and $u_{k+1}^{l+1}$. Consequently the most general form
of $F$ and $G$ is:
\[
F=F\big(k,l,u_k^l,u_k^{l+1}\big),\qquad G=G\big(k,l,u_k^l,u_{k+1}^l\big).
\]
The dependence of $F$ and $G$ upon the continuous variables $u_i^j$ is illustrated in Fig.~\ref{graph2}; together,
these functions lie on three points of the quad-graph. For this reason, we call such conservation
laws \textit{three-point
conservation laws}.
%

\noindent The three-point conservation laws can be determined directly by substituting dKdV into
\begin{gather}
F\big(k+1,l,u_{k+1}^l,u_{k+1}^{l+1}\big)-F\big(k,l,u_k^l,u_k^{l+1}\big)\nonumber\\
\qquad{}+G\big(k,l+1,u_k^{l+1},u_{k+1}^{l+1}\big)-G\big(k,l,u_k^l,u_{k+1}^l\big)=0,\label{eq15}
\end{gather}
and solving the resulting functional equation. The substitution $u_{k+1}^{l+1}=\omega$ yields
\begin{gather}
F\big(k+1,l,u_{k+1}^l,\omega\big)-
F\big(k,l,u_k^l,u_k^{l+1}\big)+G\big(k,l+1,u_k^{l+1},\omega\big)-G\big(k,l,u_k^l,u_{k+1}^l\big)=0.\label{eq10}
\end{gather}
In order to solve this functional equation we have to reduce it to a system of
 partial differential equations. To do this,
first eliminate terms that contain $\omega$, by applying each of 
the following (commuting) differential operators to (\ref{eq10}):
\[
L_1=\frac{\partial}{\partial u_k^{l+1}}-\frac{\omega_{u_k^{l+1}}}{\omega_{u_k^l}}
\frac{\partial}{\partial u_k^l},\qquad
 L_2=\frac{\partial}{\partial u_{k+1}^l}-
\frac{\omega_{u_{k+1}^l}}{\omega_{u_k^l}}\frac{\partial}{\partial u_k^l}.
\]
The operators $L_1$, $L_2$ differentiate with respect to $u_{k+1}^l$,
$u_k^{l+1}$ respectively, keeping
$\omega$ fixed. This procedure does not depend upon the form of $\omega$; 
it can be applied equally to any quad-graph equation.
In particular, for dKdV, (\ref{eq10}) is reduced to
\begin{gather}
 F_{{u_k^lu_k^l}}+ G_{{u_k^lu_k^l}}-\big(u_{k+1}^l-u_k^{l+1}\big)^2\big(F_{u_k^lu_k^{l+1}}-  G_{u_k^lu_{k+1}^{l}}\big)
 \nonumber\\
 \qquad{}- 2\big(u_{k+1}^l-u_k^{l+1}\big)\big(F_{u_k^l}+G_{u_k^l}\big)=0,\label{eq40}
\end{gather}
where $F=F(k,l, u_k^l,u_k^{l+1})$ and $G=G(k,l,u_k^l,u_{k+1}^l)$. 
Differentiating (\ref{eq40}) three times with respect to
$u_k^{l+1}$ eliminates $G$ and its derivatives, leaving the necessary condition
\begin{gather}
F_{{u_k^l}^2{u_k^{l+1}}^3}-\big(u_{k+1}^l-u_k^{l+1}\big)^2F_{u_k^l {u_k^{l+1}}^4}+
4\big(u_{k+1}^l-u_k^{l+1}\big) F_{u_k^l {u_k^{l+1}}^3}=0.\label{eq40a}
\end{gather}
This equation can be split into an overdetermined system by equating powers of $u_{k+1}^l$. 
Further information about $F$ may be found by substituting $u_{k+1}^l=\Omega$ into (\ref{eq15}).
Differentiating
\[
F\big(k+1,l,\Omega,u_{k+1}^{l+1}\big)-F\big(k,l,u_k^l,u_k^{l+1}\big)
+G\big(k,l+1,u_k^{l+1},u_{k+1}^{l+1}\big)-G\big(k,l,u_k^l,\Omega\big)=0
\]
with respect to $u_{k}^l,~u_{k+1}^{l+1}$ and keeping $\Omega$ fixed yields
\begin{gather*}
F_{{u_k^{l+1}u_k^{l+1}}}-\tilde{G}_{{u_k^{l+1}u_k^{l+1}}}
-\big(u_{k+1}^{l+1}-u_k^l\big)^2\big(F_{u_k^lu_k^{l+1}}+\tilde{G}_{u_k^{l+1}u_{k+1}^{l+1}}\big)\\
\qquad{}-2\big(u_{k+1}^{l+1}-u_k^l\big)\big(F_{u_k^l}-\tilde{G}_{u_k^{l+1}}\big)=0,
\end{gather*}
where $\tilde{G}=G(k,l+1,u_k^{l},u_{k+1}^{l+1})$.
The function $\tilde{G}$ and its derivatives are eliminated by 
differentiating three times with respect to $u_k^{l+1}$, which yields
\begin{gather}
F_{{u_k^l}^3{u_k^{l+1}}^2}-\big(u_{k+1}^{l+1}-u_k^{l}\big)^2F_{{u_k^l}^4 {u_k^{l+1}}}
+4\big(u_{k+1}^{l+1}-u_k^l\big) F_{{u_k^l}^3 {u_k^{l+1}}}=0.
\label{eq41a}
\end{gather}
The overdetermined system of partial differential equations 
(\ref{eq40a}), (\ref{eq41a}) is easily solved to obtain
\[
F=C_1u_k^lu_k^{l+1}+C_2{u_k^l}^2{u_k^{l+1}}+C_3{u_k^l}{u_k^{l+1}}^2+C_4{u_k^l}^2{u_k^{l+1}}^2+F_1+F_2,
\]
where each $C_i$ is an arbitrary function of $k$, $l$, and $F_1=F_1(k,l,u_k^{l+1})$,
$F_2=F_2(k,l,u_k^l)$ are
arbitrary functions. The term $F_2(k,l,u_k^l)$ can be 
removed (without loss of generality) by adding the trivial conservation law
\[
F_T=(\mathbf S_l-\mathbf{id})F_2,\qquad
G_T=-(\mathbf S_k-\mathbf{id})F_2,
\]
to $F$ and $G$ respectively.

So far, we have differentiated the determining equations for
 a conservation law five times; this has created a hierarchy
of functional differential equations that every three-point
 conservation law must satisfy. The unknown functions
$C_i$, $F_1$ and $G$ are found by going up the hierarchy, 
a~step at a time, to determine the constraints these equations place
on the unknown functions. As the constraints are solved sequentially, 
more and more information is gained about the
functions. At the highest stage, the determining equations are satisfied, 
and the only remaining unknowns are the constants
that multiply each conservation law. This is a simple but 
lengthy process; for brevity, we omit the details.

By this technique we have found all independent nontrivial 
three-point conservation laws for the dKdV equation; they are as 
follows\footnote{Note that the conservation laws 1 and 2 are connected to each other by a discrete symmetry of the form
$u_k^l\mapsto(-1)^{k+l}i u_k^l$; we are grateful to a referee for this observation.}:
\begin{align*}
1.~~ & F=u_k^l\big(u_k^{l+1}\big)^{2}-\big(u_k^l\big)^{2}u_k^{l+1}+u_k^l-u_k^{l+1},\\
 &G=\big(u_k^l\big)^{2}u_{k+1}^l-u_k^l\big(u_{k+1}^l\big)^{2},\\
2.~~ & F=(-1)^{k+l+1}\left\{ u_k^l\big(u_k^{l+1}\big)^{2}+\big(u_k^l\big)^{2}u_k^{l+1}-u_k^l-u_k^{l+1}\right\},\\
& G=(-1)^{k+l} \left\{ \big(u_k^l\big)^{2}u_{k+1}^l+u_k^l\big(u_{k+1}^{l}\big)^{2} \right\},\\
3.~~ & F=(-1)^{k+l+1}\left\{ \big(u_k^lu_k^{l+1}\big)^{2}-2u_k^lu_k^{l+1}+\frac{1}{2}\right\},\\
& G=(-1)^{k+l}\left\{ \big(u_k^lu_{k+1}^l\big)^{2}\right\},\\
4.~~ & F=(-1)^{k+l+1}\left\{ u_k^lu_k^{l+1}-\frac{1}{2} \right\},\\
& G=(-1)^{k+l}\left\{ u_k^lu_{k+1}^l\right\}.
\end{align*}

Three of these conservation laws depend on $k$ and $l$ explicitly. If we had chosen functions $F$ and $G$ that depended
only upon $u_i^j$ on the quad-graph (and not also upon $k$ and $l$), we would have found only the first of four
the three-point conservation laws.

\section{Five-point conservation laws}
\begin{figure}[t]
\centerline{\begin{minipage}{6.5cm}
\centerline{\includegraphics[height=6cm,width=6cm]{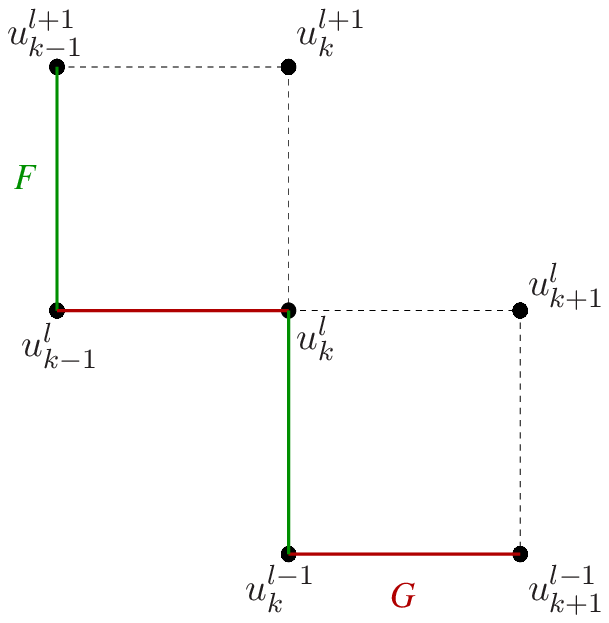}}
\vspace{-3mm}
\caption{Five-point CL.}\label{graph3}
\end{minipage}
\qquad
\begin{minipage}{6.5cm}
\centerline{\includegraphics[height=6cm,width=6cm]{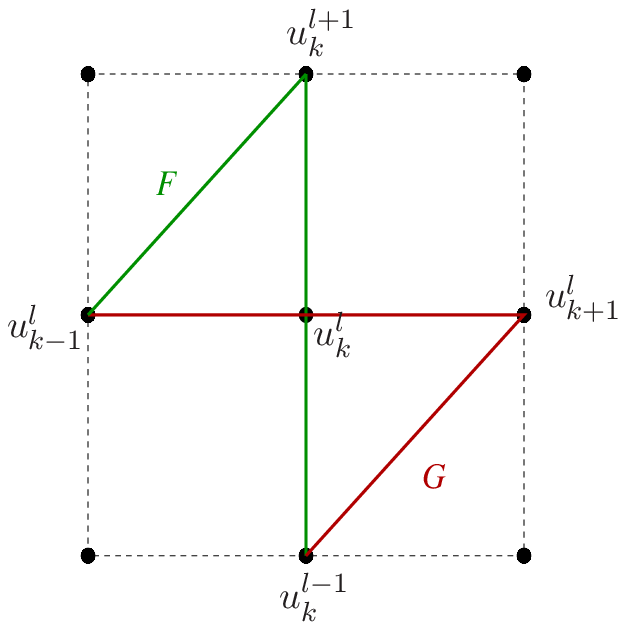}}
\vspace{-3mm}
\caption{Five-point CL.}\label{graph*}
\end{minipage}}
\end{figure}

Higher conservation laws can be found by the approach described above, but the complexity of the calculations
increases rapidly with the number of variables $u_i^j$ on which $F$ and $G$ depend. The simplest higher conservation
laws are defined on five points, as shown in Fig.~\ref{graph3}.
The functions~$F$ and~$G$ are of the form
\begin{gather*} F=F\big(k,l,u_{k-1}^{l+1},u_{k-1}^l,u_k^l,u_{k}^{l-1}\big),\qquad
G=G\big(k,l,u_{k-1}^l,u_k^l,u_{k}^{l-1},u_{k+1}^{l-1}\big).
\end{gather*}
The points lie in a pair of quad-graphs with a single common point.
We seek conservation laws that contain points from both quad-graphs and cannot be reduced to the conservation
laws from the previous section (modulo trivial conservation laws).

The determining equation for the five-point conservation laws is
\begin{gather*} F(k+1,l,u_k^{l+1},u_{k}^l,u_{k+1}^l,u_{k+1}^{l-1})
-F(k,l,u_{k-1}^{l+1},u_{k-1}^l,u_k^l,u_{k}^{l-1})\\
\qquad{}+G(k,l+1,u_{k-1}^{l+1},u_k^{l+1},u_{k}^{l},u_{k+1}^l)
-G(k,l,u_{k-1}^l,u_k^l,u_{k}^{l-1},u_{k+1}^{l-1})=0.
\end{gather*}
Shifted versions of the dKdV equation are used to 
eliminate $u_{k+1}^l$, $u_k^{l+1}$ in favour of the variab\-les~$u_i^j$ at the five
points that lie on the `step' shown in bold in Fig.~\ref{graph3}. The determining equation can be solved by the same
technique which is described in the previous section. This is a very lengthy
calculation, so we state the results only.

The extra five-point conservation laws of dKdV equation are
\begin{align*}
5.~~ & F=\ln  \big( u_k^l-u_{k-1}^{l+1}\big),\\
& G=\ln  \left(   \frac{1}{u_{k+1}^{l-1}-u_k^l}-u_{k-1}^l+u_k^{l-1} \right),\\
6.~~ & F=\ln  \left(  \frac{1}{u_k^{l-1}-u_{k-1}^{l}} -u_k^l+u_{k-1}^{l+1} \right),\\
& G=\ln\big(u_{k-1}^l-u_k^{l-1}\big),\\
7.~~ & F=l\ln  \left( u_k^l-u_{k-1}^{l+1}+ \frac{1}{u_{k-1}^l-u_k^{l-1}}  \right) -( k-1 )
\ln\big( u_{k-1}^{l+1}-u_k^l\big),\\
&  G=l\ln \big( u_{k-1}^l-u_k^{l-1} \big)-k\ln  \left(  \frac{1}{ u_{k+1}^{l-1}-u_k^l}-u_{k-1}^l+u_k^{l-1} \right).
\end{align*}

These conservation laws are irrational, so it is clear that they cannot be reduced to the polynomial conservation
laws in the previous section. The five-point conservation laws can be simplified somewhat with the substitutions
\begin{gather*}
u_{k-1}^{l+1}=u_k^l+\frac{1}{u_{k-1}^l-u_k^{l+1}},\qquad u_{k+1}^{l-1}=u_k^l+\frac{1}{u_k^{l-1}-u_{k+1}^l}.
\end{gather*}
These substitutions move the five-point conservation laws onto the cross shown in Fig.~\ref{graph*}.
%
The functions $F$ and $G$ are now of the form
\begin{gather*} F=F'\big(k,l,u_{k-1}^l,u_k^{l-1},u_k^l,u_k^{l+1}\big),\qquad
G=G'\big(k,l,u_{k-1}^l,u_k^{l-1},u_k^l,u_{k+1}^l\big).
\end{gather*}
Specifically, the five-point conservation laws have the following components:
\begin{align*}
5.~~ & F'=-\ln \big( u_k^{l+1}-u_{k-1}^l\big),\\
&  G'=\ln\big(u_{k+1}^l-u_{k-1}^l\big),\\
6.~~ & F'=\ln\left({\frac {u_k^{l-1}-u_k^{l+1}}{\big(u_k^{l+1}-u_{k-1}^l\big)\big(u_{k-1}^l-u_k^{l-1}\big)}}\right),\\
&  G'=\ln\big(u_{k-1}^l-u_k^{l-1}\big),\\
7.~~ &  F'=l\ln\left({\frac{u_k^{l+1}-u_k^{l-1}}{\big(u_k^{l+1}-u_{k-1}^l\big)\big(u_{k-1}^l-u_k^{l-1}\big)}}\right)
+(k-1)\ln\big(u_{k-1}^l-u_k^{l+1}\big),\\
& G'=l\ln\big(u_{k-1}^l-u_k^{l-1}\big)-k\ln\big(u_{k+1}^l-u_{k-1}^l\big).
\end{align*}
Surprisingly, none of them do depend upon $u_k^l$. We will describe the circumstances under which this
occurs for other quad-graph equations in a separate paper.

\section{Conclusion and outlook}
\begin{figure}[t]
\centerline{\includegraphics[height=7cm,width=7cm]{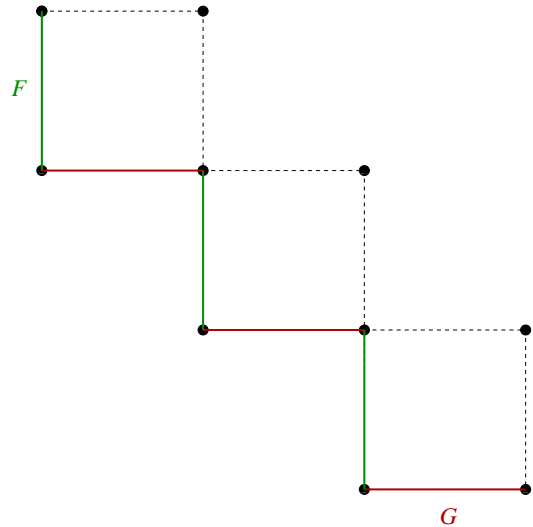}}
\vspace{-3mm}

\caption{Seven-point CL.}\label{graph4}
\end{figure}

The main results of this paper are as follows.
\begin{itemize}
\itemsep=0pt
\item New conservation laws for the dKdV equation have been found.
\item These include higher-order conservation laws, which are irrational.
\item We have improved the effectiveness of Hydon's direct method for constructing conservation laws \cite{Hy0}.
\end{itemize}
In principle, the same method can be used to 
construct conservation laws with seven or more points (Fig.~\ref{graph4}).
However, the calculations become extremely complex, placing heavy demands on even the most sophisticated computer algebra
systems. We are currently working to improve the efficiency of the method still further, but it is unlikely that it will
be possible to determine conservation laws of very high order directly. However, we are developing ways
of combining direct and indirect methods to achieve this aim, as will be reported elsewhere.

\LastPageEnding

\end{document}